\documentclass[12pt]{article}
\usepackage{amssymb}
\usepackage{amsmath, epsfig, color}

\newcommand{\be}{\begin{equation}}
\newcommand{\en}{\end{equation}}

\renewcommand{\vec}[1]{\boldsymbol{#1}}

\newcommand{\demi}{\textstyle{\frac{1}{2}}}

\begin{document}

\title{Third- and fourth-order constants  of incompressible soft solids \\ and the acousto-elastic effect}


\author{%
Michel Destrade$^a$, Michael D. Gilchrist$^a$, G. Saccomandi$^b$\\[12pt]
School of Electrical, Electronic, and Mechanical Engineering, \\
University College Dublin, \\
Belfield, Dublin 4, Ireland.\\[12pt]
Universit\`{a} degli Studi di Perugia,   \\   06125 Perugia, Italy.}

\date{}
\maketitle

\begin{abstract}

Acousto-elasticity is concerned with the propagation of small-am\-pli\-tu\-de waves in deformed solids.
Results previously established for the incremental elastodynamics of exact non-linear elasticity are useful for the determination of third- and fourth-order elastic constants, especially in the case of incompressible isotropic soft solids, where the expressions are particularly simple. 
Specifically, it is simply a matter of expanding the expression for $\rho v^2$, where $\rho$ is the mass density and $v$ the wave speed, in terms of the elongation $e$ of a block subject to a uniaxial tension. 
The analysis shows that in the resulting expression: $\rho v^2 = a + b e + c e^2$, say, $a$ depends linearly on $\mu$; $b$ on $\mu$ and $A$; and $c$ on $\mu$, $A$, and $D$, the respective second-, third, and fourth-order constants of incompressible elasticity, for bulk shear waves and for surface waves.

\end{abstract}

\newpage


\section{INTRODUCTION}


Recent years have witnessed a surge of interest in the acoustics of \emph{incompressible} soft solids, with the long-term goal of mastering all aspects of ultrasonic wave propagation in biological soft tissues. 
In the wake of a 2004 article by Hamilton et al.\cite{HaIZ04}, establishing that the general strain energy $W$ of \emph{fourth-order incompressible elasticity} involves only three elastic constants,
\be \label{incomp4}
W = \mu \; \text{tr}\left(\vec{E}^2\right)  + \frac{A}{3} \;\text{tr}\left(\vec{E}^3\right)   + D \; \left(\text{tr} (\vec{E}^2)\right)^2,
\en
where $\vec{E}$ is the Green strain tensor and $\mu$, $A$, and $D$ are second, third-, and fourth-order elasticity constants, respectively, at least 16 articles\cite{HaIZ04B, ZaIHM05, HaIZ07, Genn07, Reni07, JCGB07, OsSu07, DeSa08, WHIZ08, Woch08, Reni08b, Reni08a, Ostro08, ZaIH09, Doma09, DeJS09} have studied the dynamics of those solids.

Although it is of course important to make progress in this area, because of obvious repercussions for elastography techniques, it is also important to acknowledge, and \emph{to use}, related and previously established results, such as those coming from the literature on incremental (small-on-large) waves in homogeneously deformed materials, because they give direct access to a simple way to evaluate $\mu$, $A$, and $D$.
In fact, the acousto-elasticity of non-linearly elastic incompressible solids has a long and strong history, dating back to the seminal works of R.S. Rivlin, M.A. Hayes, M.A. Biot, and many others.
For instance, Rivlin and Saunders\cite{RiSa51} find in 1951 that only two constants are needed to encompass third-order incompressible elasticity (where $W$ is given by Eq.\eqref{incomp4} written at $D =0$), and Ogden\cite{Ogde74} shows, as early as 1974, that only three constants are required at fourth-order, as in, and preceding, Eq.\eqref{incomp4}.

The present paper shows that acoustic plane waves can be used to measure the third- and fourth-order elastic constants of incompressible soft solids. 
This goal can be reached using \emph{non-linear} waves, but these are difficult to generate and to detect experimentally in solids.
Moreover, although it might be possible to observe them in liquid-like solids\cite{Reni08b} (such as gels, phantoms, agar, etc.), it seems harder to imagine that they could travel in biological soft tissues without causing damage at the cellular  level, by virtue of them being of large amplitude (see Mironov et al.\cite{Miro09} for experimental evidence of such damage).
On the other hand, linear, or rather \emph{linearized}, acoustic waves can also give the non-linear elastic constants, through their coupling with a pre-deformation: this is the so-called \emph{acousto-elastic effect}.

An abundance of results exists in the literature on incremental dynamics in exact non-linear elasticity, usually giving the speed of a small-amplitude wave propagating in \emph{any} incompressible solid, subject to \emph{any} homogeneous pre-deformation.
It is thus a simple matter to specialize these results to the strain energy density of Eq.\eqref{incomp4}, and to a uni-axial tension  yielding a stretch $\lambda = 1 + e$, say, where $e$ is the \emph{measurable} elongation.  

When an \emph{isotropic} soft solid is subjected to a uni-axial tension ($e>0$) or compression ($e<0$), it exhibits \emph{strain-induced anisotropy}.
The axes of anisotropy are the so-called \emph{principal axes}, aligned with the principal axes of the Cauchy-Green strain ellipsoid. 
The squared wave speed of an infinitesimal wave traveling in an incompressible solid subject to a small-magnitude uni-axial pre-deformation depends (at first) linearly on the elongation $e$, and the coefficients of linear interpolation are linear combinations of $\mu$ and $A$.
Conversely, measuring the wave speed gives access to the values of $\mu$ and $A$.
In order to access $D$, the fourth-order constant of \eqref{incomp4}, the characteristics of a small-amplitude wave must be coupled to a small-but-finite uni-axial pre-deformation, i.e. yielding $v^2$ as a quadratic function of $e$.
In other words, this paper aims at finding explicitly the coefficients $a$, $b$, and $c$ in the expansion
\be \label{expan}
\rho v^2  = a + b e + c e^2,
\en
where $\rho$ is the mass density. 
For bulk homogeneous waves (Section III), we find of course that $a=\mu$, the initial shear modulus, giving the linear wave speed $\sqrt{\mu/\rho}$.
Similarly, we find that $a = 0.9126 \mu$ for surface waves (Section IV), consistent with Lord Rayleigh's result\cite{Rayl85} in linear isotropic incompressible solids.
For both types of waves, we find that $b$ depends linearly on $\mu$ and $A$, and that $c$ depends linearly on $\mu$, $A$, and $D$.


\section{PRE-STRESS, PRE-DEFORMATION,\\ ELASTIC MODULI}


Take a parallepipedic sample of a soft, isotropic, incompressible solid
and subject it to a uni-axial  tension or compression, and call $\lambda$ the resulting stretch in the direction of elongation.
It is a simple and common exercise of non-linear (exact) elasticity to show that the corresponding principal stretches are:
\be \label{stretches}
\lambda_1 = \lambda, \qquad
\lambda_2 = \lambda^{-\demi}, \qquad
\lambda_3 = \lambda^{-\demi},
\en
showing that the deformation is equi-biaxial.
We call $x_1$ the axis along the direction of elongation, and $x_2$, $x_3$, two orthogonal axes in the plane normal to that direction, such that the faces of the deformed block are parallel to the ($x_1,x_2$)-, ($x_2, x_3$)-, and ($x_3, x_1$)-planes.

With respect to incremental wave propagation, some quantities, $\gamma_{ij}$, $\beta_{ij}$  ($i , j = 1,2,3$), play a privileged role, akin to that of elastic moduli. 
They are defined in general by\cite{DoOg90, DOPR05}
\begin{align}
& \gamma_{ij} = (\lambda_i W_i - \lambda_j W_j)\lambda_i^2/(\lambda_i^2 - \lambda_j^2) \ne \gamma_{ji}, 
\notag \\ 
& \beta_{ij}  = \demi \left(\lambda_i^{2} W_{ii} +  \lambda_j^2 W_{jj}\right) - \lambda_i \lambda_j W_{ij}
-  (\lambda_j W_i - \lambda_i W_j)\lambda_i \lambda_j/(\lambda_i^2 - \lambda_j^2) = \beta_{ji},
\end{align}
when $\lambda_i \ne \lambda_j$, and by 
\begin{align}
& \gamma_{ij} = \demi \left(\lambda_i^2 W_{ii} - \lambda_i\lambda_j W_{ij} + \lambda_i W_i\right) \ne \gamma_{ji}, 
\notag \\ 
& \beta_{ij}  = \demi \left(\lambda_j^{2} W_{jj} - \lambda_i \lambda_j W_{jj} - \lambda_i W_i \right)  = \beta_{ji},
\end{align}
when $\lambda_i = \lambda_j$.
Here $W_i \equiv \partial W / \partial \lambda_i$, $W_{ij} \equiv \partial^2 W/\partial\lambda_i\partial \lambda_j$,  and $i \ne j$ (also, there are no sums on repeated indices).
We recall that the eigenvalues $E_i$ of the Green strain tensor are related to the $\lambda_i$ by $E_i = (\lambda_i^2-1)/2$.
It is thus easy to write $W$ of Eq.\eqref{incomp4} in terms of the $\lambda_i$ and then to relate the results to Eq.\eqref{stretches}.
Finally, the results can easily be expanded in terms of the \emph{elongation} $e = \lambda -1$.
Here we find that 
\begin{align} \label{alpha}
& \gamma_{12} = \gamma_{13}  
= \mu   + \left(3\mu +A/4 \right)e + \left(5\mu + 7A/4 + 3D\right)e^2 + \mathcal{O}(e^3),
 \notag \\
& \gamma_{21} = \gamma_{31} 
= \mu  +(A/4) e + (2\mu + A + 3D)e^2 + \mathcal{O}(e^3),
 \notag \\
& \beta_{12} = \beta_{13} 
= \mu + \left(3\mu/2 + A/4\right) e +\left(25\mu/2 + 47A/8 + 12D\right) e^2 + \mathcal{O}(e^3),
 \notag \\
& \gamma_{23} = \gamma_{32} = \beta_{32} 
= \mu - \left(3\mu + A/2\right) e +\left( 5\mu + 7A/4 + 3D \right) e^2 + \mathcal{O}(e^3),
 \end{align}
and we do not need to specify the higher-order terms for our purpose, which is to establish relationships in the form of Eq.\eqref{expan}.
Notice that 
 \begin{multline}
 2 \beta_{12} - \gamma_{12} - \gamma_{21} =
 2 \beta_{13} - \gamma_{13} - \gamma_{31} = 
 12 \mu \nu  e^2 +  \mathcal{O}(e^3), \\
 \quad \text{where} \quad 
 \nu \equiv \frac{3}{2}\left(1 + \frac{A/2 + D}{\mu} \right)
 \end{multline}
 is the \emph{coefficient of non-linearity} for linearly polarized plane shear waves, obtained by Hamilton et al.\cite{HaIZ04}


\section{BODY WAVES}


In a deformed \emph{incompressible} solid, two infinitesimal transverse (shear) homogeneous plane waves may propagate (but no longitudinal homogeneous plane wave), which are linearly polarized along mutually orthogonal directions. 
Here we consider the propagation of homogeneous plane waves in the ($x_1,x_2$)-, ($x_2,x_3$)-, and ($x_3, x_1$)-planes, the so-called \emph{principal planes}.

\begin{figure}
\begin{center}
\includegraphics*[width=10cm]{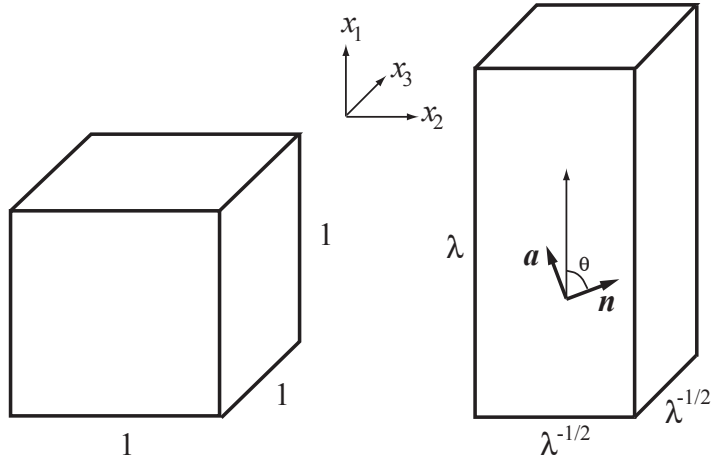}
\end{center}
\caption{Deformation of a unit cuboid in an initially isotropic, incompressible solid, elongated by the amount $e=\lambda-1$ (in this sketch, the cube is in tension and $e>0$). The shear bulk waves under study propagate in either of the three principal planes (here, the ($x_1, x_2$)-plane), in the direction of the unit vector $\vec{n}$, at an angle $\theta$ with a principal direction (here, the $x_1$-direction). One wave is polarized in the principal plane along the unit vector $\vec{a}$ orthogonal to $\vec{n}$, the other wave is polarized along $\vec{a} \times \vec{n}$, the normal to the principal plane. 
}
\label{fig_1}
\end{figure}

\begin{figure}
\begin{center}
\includegraphics*[width=10cm]{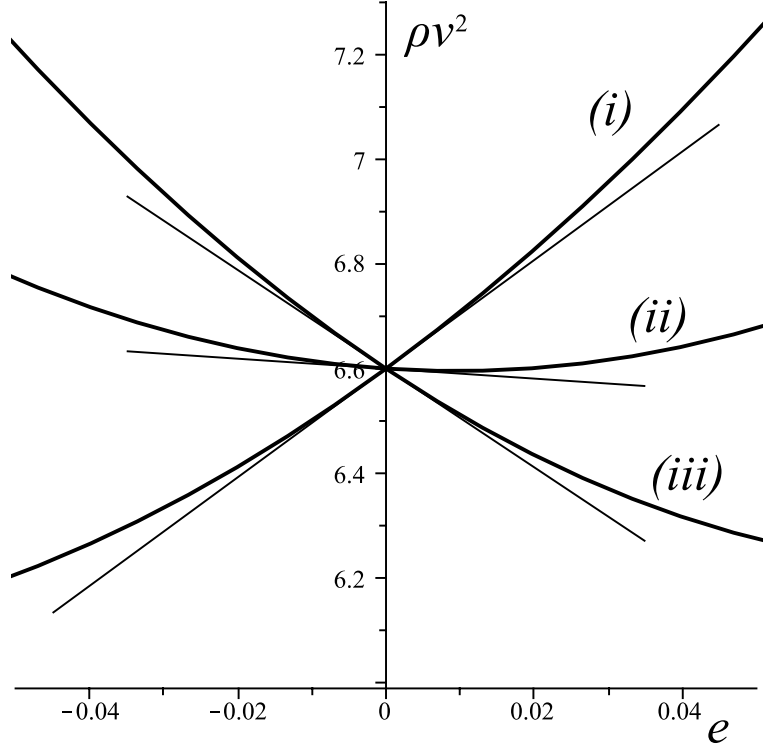}
\end{center}
\caption{
Shear bulk waves propagating in a uni-axially deformed incompressible block: $\rho v^2$ (in kPa) against the elongation $e$ in the case of waves traveling \emph{(i)} in the $x_1$-direction (then, all shear waves travel with speed given by Eq.(12), irrespective of the direction of polarization); \emph{(ii)} in any direction in the ($x_1, x_2$)-plane or in the ($x_1, x_3$)-plane, with transverse polarization in that plane, and speed given by Eq.(13);  \emph{(iii)} in any direction in the ($x_1, x_2$)-plane or in the ($x_1, x_3$)-plane, with polarization normal to that plane, and speed given by Eq.(14). The thin straight lines correspond to the linear part of the $\rho v^2$-$e$  curves. 
Here the initial shear modulus is $\mu =6.6$ kPa, the third-order Landau constant is $A=-37.7$ kPa, and the coefficient of non-linearity is $\nu = 3.5$, in line with the experimental results of Renier et al.\cite{Reni07}. 
}
\label{fig_2}
\end{figure}

First consider that the waves travel in the ($x_1, x_2$)-plane. 
In that plane, call $\vec{n}$ the unit vector in the direction of propagation and call $\vec{a}$ the unit vector orthogonal to $\vec{n}$, see Figure \ref{fig_1}.
Based on previous studies (see, for example, Ogden\cite{Ogde07}), we find that one of the two shear bulk waves is polarized along $\vec{a}$, and travels with speed $v_a$, given in terms of the elastic moduli by the following \emph{secular equation}, 
\be \label{bulk1}
\rho v_a^2 = (\gamma_{12} + \gamma_{21} - 2\beta_{12})\cos^4\theta + 2(\beta_{12} - \gamma_{21})\cos^2\theta + \gamma_{21},
\en
in general, where $\theta$ is the angle between $\vec{n}$, the direction of propagation and $x_1$, the direction of the uni-axial pre-deformation (both the angle and the phase speed are evaluated in the deformed configuration). 
Using the expressions Eq.\eqref{alpha} for $\gamma_{12}$, $\beta_{12}$, and $\gamma_{21}$, we arrive at a relationship in the form of Eq.\eqref{expan}, where
\begin{multline}
a = \mu, \qquad
b= 3\mu \cos^2\theta + A/4, \\
c = 2\mu + A + 3D + (21\mu + 39A/4 + 18D)\cos^2\theta - 12\mu \nu \cos^4\theta.
\end{multline}
The relationship provides an obvious way to determine $\mu$, $A$, and $D$ experimentally, either by fixing $e$ and varying $\theta$, or vice-versa, see the example presented in Fig.\ref{fig_2}.
The other bulk shear wave is polarized along $x_3$ (i.e. along $\vec{a} \times \vec{n}$), and its wave speed $v_b$ is given by
\be \label{bulk2}
\rho v_b^2 = \gamma_{13} \cos^2\theta + \gamma_{23}\sin^2\theta,
\en
in general or, using the expressions Eq.\eqref{alpha} for $\gamma_{13}$ and $\gamma_{23}$, also by a relationship in the form of Eq.\eqref{expan}, where now
\be 
a = \mu, \qquad
b= 3(\mu + A/8)\cos(2\theta) - A/8, \qquad
c = 5\mu + 7A/4 + 3D.
\en
Notice here how the quadratic coefficient in the $\rho v^2$--$e$ relation is independent of the angle of propagation.
Note also that in the case of \emph{principal wave propagation along} $x_1$, the direction of elongation, $\theta=0$ and the two shear wave speeds coincide:
\be \label{ppl}
\rho v_a^2  = \rho v_b^2 = \mu + \left(3\mu + A/4 \right)e + \left( 5\mu + 7A/4 + 3D \right)e^2,
\en
showing that $x_1$ is an acoustic axis, along which circularly-polarized waves may propagate.

Next, consider that the waves travel in the ($x_1, x_3$)-plane. 
Again, call $\theta$ the angle between the direction of propagation and $x_1$, the direction of the uni-axial pre-deformation. 
Then we find that, owing to the equalities in Eq.\eqref{alpha}, the secular equations are exactly the same as for propagation in the ($x_1, x_2$)-plane. 
This is consistent with the transversely isotropic character of the strain-induced anisotropy.

Finally, consider waves traveling in the ($x_2, x_3$)-plane.
We call $\theta$ the angle between the direction of propagation and $x_2$, so that $\vec{n} = [0, \cos \theta, - \sin \theta]^T$ and $\vec{a} = [0, \sin \theta, \cos \theta]^T$. 
Then the secular equation for the shear wave polarized along $\vec{a}$ is Eq.\eqref{bulk1} where the indices 12 and 21 are replaced by 23 and 32, respectively. 
However, because $\gamma_{23} = \gamma_{32} = \beta_{32}$, the dependence on $\theta$ vanishes, and the wave propagates in an isotropic manner in that plane.
Its speed is given by 
\be \label{iso1}
\rho v_a^2 =  \gamma_{23} 
= \mu - \left(3\mu + A/2\right) e +\left( 5\mu + 7A/4 + 3D \right) e^2.
\en
Similarly, the secular equation for the shear wave polarized along $\vec{n} \times \vec{a}$, i.e. along $x_1$, the direction of elongation, is Eq.\eqref{bulk2}, where the indices 13 and 23 are replaced by 21 and 31, respectively. 
However, because $\gamma_{21} = \gamma_{31}$, the dependence on $\theta$ vanishes and that wave also propagates in an isotropic manner in the ($x_2, x_3$)-plane.
This observation is consistent with the transversely isotropic character of the induced anisotropy.
In that latter case, the speed is given by 
\be \label{iso2}
\rho v_b^2 =  \gamma_{21} 
= \mu  +(A/4) e + \left( 2\mu + A + 3D \right)e^2 .
\en

Now we make the link with the theoretical results of Gennisson et al.\cite{Genn07}, who derived the \emph{linear} acousto-elastic dependence of the squared wave speed on the uni-axial stress $\sigma$ (say), in the cases of principal wave propagation.
To establish this connection, it suffices to keep the linear part of the $\rho v^2$--$e$ relations Eqs.\eqref{ppl}, \eqref{iso1}, and \eqref{iso2}, and to recall that for incompressible solids $e = -\sigma/(3\mu) + \ldots$, to recover the following expansions\cite{Genn07},
\be 
\rho v^2 = \mu -\sigma \left(1 + \dfrac{A}{12 \mu}\right), \qquad \mu + \sigma\left( 1 + \dfrac{A}{6\mu} \right),
\qquad  \mu - \sigma\left( \dfrac{A}{12\mu} \right),
\en
respectively. 

We conclude this section with an example taken from experimental investigations. 
Renier et al.\cite{Reni07} use a small elongation (compression) to deduce a linear $\rho v^2$--$\sigma$ relationship from experimental data on Agar-Gelatin based phantoms: 
they conclude that for their 5\% Gelatin sample, $\mu \simeq 6.6$ kPa and $A \simeq -37.7$ kPa. 
Then they use finite-amplitude shear waves to estimate the coefficient of non-linearity: 
they find $\nu \simeq 3.5$ (see also the combination of Ref.\cite{Genn07} and Ref.\cite{Reni08b}.)
Eqs.\eqref{bulk1}-\eqref{iso2} provide an alternative mean of determining $\nu$ for that sample, using \emph{infinitesimal}, non-destructive, shear waves instead of finite-amplitude shear waves.
As seen clearly on Fig.\ref{fig_2}, the (theoretical) linear part of the $\rho v^2$--$e$ is confined to $\pm 1 \%$ stretches, approximatively,  and it suffices to elongate  the block by a few percent to perceive the influence of the fourth-order elasticity constant.


\section{SURFACE WAVES}


Dowaikh and Ogden\cite{DoOg90} established the secular equation for a surface acoustic wave propagating in a principal direction of a deformed incompressible solid (see also Hayes and Rivlin\cite{HaRi61}).
Here, for a wave traveling in the direction of elongation $x_1$, with attenuation in the $x_2$-direction, it reads in general as
\be \label{dowaikh}
\eta^3 + \eta^2 +  (2\beta_{12} + 2\gamma_{21} - \gamma_{12})\eta/\gamma_{21} - 1 = 0,
\qquad
\text{where}
\qquad
 \eta \equiv \sqrt{(\gamma_{12} - \rho v^2)/\gamma_{21}}.
\en
With $\gamma_{12}$, $\beta_{12}$, and $\gamma_{21}$ given by  Eq.\eqref{alpha}, the secular equation Eq.\eqref{dowaikh}$_1$ is
\be \label{f}
f(\eta, e^2) \equiv \eta^3 + \eta^2 +  (3 + 12\mu \nu e^2)\eta - 1 = 0.
\en
Clearly now, $\eta$ is of the form $\eta = \eta_0 + \eta_2 e^2$ plus higher-order terms, and no term in $e^1$. 
Here, $\eta_0=0.2956$ is the unique positive real root of the cubic 
$\eta^3 + \eta^2 +  3\eta - 1 = 0$. 
To find $\eta_2$, we expand Eq.\eqref{f}, and conclude that 
\be
\eta_2 = - \dfrac{\dfrac{ \partial f}{\partial(e^2)}(\eta_0,0)}{\dfrac{\partial f}{\partial \eta}(\eta_0, 0)} = -\dfrac{12 \eta_0}{ (3\eta_0^2 + 2\eta_0 + 3)} \mu \nu = 
0.9205 \mu \nu.
\en
It follows from this and Eq.\eqref{dowaikh}$_2$ that $\rho v^2 = \gamma_{12} - \gamma_{21}\eta_0(\eta_0+2\eta_2e^2)$, or
\be
 \rho v^2  = 0.9126 \mu + (3 \mu + 0.9126A/4)e + (5.642 \mu + 2.071A + 3.554 D)e^2.
 \en 
Here, the numbers 3 and 4 are exact, whilst 0.9126 is $x_R \equiv 1-\eta_0^2$, the celebrated root of  Rayleigh's  cubic \cite{Rayl85}:
$x^3 - 8x^2 + 24x -16=0$, for surface waves in linear, isotropic, incompressible solids.

The principal Rayleigh wave propagating in the $x_1$-direction, with attenuation in the $x_3$-direction has the same secular equation, due to the equalities $\gamma_{13} = \gamma_{12}$, $\gamma_{23}=\gamma_{13}$, and $\beta_{13}=\beta_{12}$.

To find the speed of a principal surface wave propagating in the $x_2$-direction, with attenuation in the $x_1$-direction (or equivalently, propagating in the $x_3$-direction, with attenuation in the $x_1$-direction), we swap the roles played by $\gamma_{12}$ and $\gamma_{21}$ in Eq.\eqref{dowaikh}. 
In that case we conclude that
 \be
 \rho v^2 = 0.9126 \mu - (0.2621 \mu - 0.2281 A) e + (2.379 \mu + 1.255A + 3.554 D)e^2.
 \en 
 

Finally, the speed of surface waves propagating on the faces parallel to the ($x_2, x_3$)-plane cannot be measured, because those faces are clamped by the uni-axial tension device.







\section{CONCLUDING REMARKS}


There are two different paths to the study of non-linear elastic effects. 
The \emph{theory of exact non-linear elasticity} aims at describing large deformations, with no cap on the magnitude of the mechanical fields. 
For instance, it tries to find a strain energy density $W$ which can describe the mechanics governing the full extension of a rubber string, \emph{i.e.} up to stretches of 200-800\%. 
The constitutive parameters of $W$ are determined from experimental data through a global minimization procedure, by non-linear curve fitting over the whole stretch range. 
However it can be argued that the search for a good model encompassing the whole non-linear mechanical behavior of elastomers and rubber-like materials is a scientific chimera: so far, many forms of strain energies have been proposed, but none has given an accurate description of real-world materials for all deformation fields. 
By contrast, the \emph{theory of weakly non-linear elasticity} takes a step-by-step approach to non-linearity, and determines constitutive parameters one by one, as they appear successively at the onset of non-linear effects. 
This goal is quite simple to achieve experimentally. 
However, the mathematics involved in the resolution of increasingly non-linear systems of equations are quite intricate and sometimes, unnecessary, because exact results already exist in exact non-linear elasticity, from which it is easy to deduce weakly non-linear expansions at the required order.

This paper aimed at uncovering one of these overlaps of the two theories, specifically the acousto-elastic effect in incompressible elasticity. 
For example (\S III), it is possible to measure the speed $v$ of a bulk wave propagating in an elongated block. 
The squared wave speed is linearly related to the elongation $e$, as long as $e$ is small.
The coefficients of linear correlation give the second- and third-order elasticity constants. 
When $e$ is increased, the $v^2$ -- $e$ relation becomes quadratic, and the coefficient of $e^2$ gives the fourth-order elastic constant. 
Similar results are found using surface waves (\S IV).

These connections are fruitful for isotropic non-linear elastic solids, and others can easily be established in, for example, the cases of reflected and refracted waves, torsional waves, plate waves, etc. 
They provide useful analytical benchmarks for the experimental determination of third- and fourth-order elastic constants of soft gels and silicones. 
For biological soft tissues however, \emph{anisotropy} must be included into the analysis, because of the presence of collagen fiber bundles. 
This inclusion can have dramatic consequences with respect to small-on-large theory\cite{DeGM09}.
It also raises rapidly the number of elastic constants to be determined. 
For instance, the introduction of just one family of parallel fibers in an incompressible isotropic matrix raises the number of constants from 2 to 7 in third-order elasticity, and from 3 to 13 in fourth-order elasticity\cite{DeGO10}.

Finally, we mention experimental issues. 
Obviously, generating, controlling, and measuring a \emph{nonlinear} (finite-amplitude) shear wave is a delicate affair, subject to potential problems arising from beam diffraction, temperature changes due to wave intensity, microstructural damage, etc.
By contrast, \emph{infinitesimal} (small-amplitude) acoustic waves are routinely produced by bulk (BAW) or surface (SAW) acoustic wave transducers.
This paper shows that the coefficient of nonlinearity for shear waves can be obtained thanks to the coupling of an infinitesimal BAW or a SAW with the pre-strain. 
Acousto-elasticity itself is a relatively old branch of physical acoustics, that can be dated back to the experiments of Hughes and Kelly\cite{HuKe53} in 1953, and is now integrated in standard Handbooks, see for example Refs.\cite{PaSF84, KiSa01}.
It is employed to determine experimentally third-order elasticity constants of ordinary elastic materials within their elastic limit. 
Fourth-order constants cannot be measured in general for metals, crystals, rocks, and other ``stiff'' elastic solids, because of plastic yield, thermal effects, phase transformation, etc. 
For ``soft'' elastic solids, however, the elastic limit is far beyond 10\% (up to hundreds of percent for rubbers).
It follows that if the acousto-elastic effect can be measured in these solids for infinitesimal pre-strains, then it can also be detected, in exactly the same experimental conditions, for slightly larger pre-strains. 
As Fig.\ref{fig_2} suggests, a pre-strain of a couple percent is sufficient to pick up the fourth-order elasticity constant.


\end{document}